\documentclass[letterpaper]{jpconf}
\usepackage{graphicx}
\begin{document}
\title{Relevance of Ion-Channeling for Direct DM Detection\footnote{Talk given at  the TAUP 2009 conference, Rome, Italy, July 1-5 2009}}

\author{Graciela B. Gelmini}

\address{Department of Physics and Astronomy, UCLA, 475 Portola Plaza, Los
  Angeles, CA 90095, USA- E-mail: gelmini@physics.ucla.edu}

\begin{abstract}

The channeling   of the recoiling nucleus in crystalline detectors after a WIMP collision would produce a larger scintillation  or ionization signal
in direct detection experiments  than otherwise expected.   I present estimates  of the importance of this effect for the total direct detection rate and the daily modulation of the signal using analytic models produced in the 1960's and 70's  to describe the effects of channeling and blocking in crystals.
\end{abstract}

Channeling and blocking effects in crystals refer to the orientation dependence of positively charged ion penetration in crystals~\cite{Gemmell:1974ub}. In the ``channelling effect" ions incident upon a crystal along symmetry axis and planes suffer a series of small-angle scattering  that maintain them in the open``channels"  in between the rows or planes of lattice atoms and thus penetrate much further into the crystal than in other directions. Channeled incident ions do not get close to lattice sites, where they would experience large-angle scatterings. The ``blocking effect"  consists in a reduction of the flux of ions originating in lattice sites along symmetry axis and planes, creating what is called a ``blocking dip" in the ions exiting from a thin enough crystal as function of the angle with respect to the symmetry axis or plane. These effects were first observed in the 1960's and  are currently extensively used in crystallography,  in the study  of lattice disorder, ion implantation, surfaces, interfaces and epitaxial layers, short nuclear lifetimes etc.

The channeling effect in NaI (Tl) was first observed in 1973 by Altman {\it et al.}~\cite{Altman} who proved  that channeled ions produce more scintillation light. The reason is that  they loose most of their energy via electronic stopping rather than nuclear stopping. In 2007 Drobyshevski~\cite{Drobyshevski:2007zj} and the 
DAMA collaboration~\cite{Bernabei:2007hw} pointed out the
relevance of this effect  for direct detection experiments using NaI (Tl) crystals: when  Na or I ions recoiling after a collision with a WIMP move along crystal axes and planes, their quenching factor is approximately $Q=1$ instead of $Q_I=0.09$ and $Q_{Na}=0.3$. In 2008 Avignone, Creswick and  Nussinov~\cite{Avignone:2008cw} raised the prospect of a daily modulation of the DM signal in direct detection  due to channeling. In fact Earth's daily rotation makes the WIMP wind change direction with respect to the crystal, which  produces a daily modulation in the measured recoil energy (equivalent to a modulation of the factor $Q$). 

My collaborators, Nassim  Bozorgnia and Paolo Gondolo, and I~\cite{BGG} are using classical analytic models of channeling  developed in the 1960's and 70's, in particular Lindhard's model~\cite{Lindhard:1965}
 to calculate the fraction of channeled ions as function of their energy and  the expected daily modulation amplitudes due to channelling in NaI, Si and Ge. The results presented here are preliminary.

We use the continuum string and plane model, in which the screened Thomas-Fermi potential is averaged over a direction parallel to a row or a plane. This averaged potential $U$ is considered to be uniformly smeared out along the row or plane of atoms, what is a good approximation if the propagating ion interacts with many lattice atoms in the row or plane by a correlated series of many consecutive glancing collisions with lattice atoms. Just one row, or one plane, is considered.  We use the ``Radon transform" of the WIMP velocity distribution (a truncated Maxwellian) to write the  ion recoil momentum spectrum in terms of the average velocity of the WIMP's with respect to the laboratory, the velocity dispersion  and escape speed~\cite{Gondolo:2002}. We orient the crystal with respect to the galaxy and use the 
Hierarchical Equal Area isoLatitude Pixelization (HEALPix) method~\cite{HEALPix:2005} to compute the integrals over the recoil direction. The HEALPix method uses an algorithm to deal with the pixelization of data on a sphere (used e.g. by cosmic microwave background experiments). It is useful to compute integrals over direction by dividing the surface of a sphere into many pixels, computing the integrand at each pixel (i.e. each direction, see Fig.~1) and finally summing up the value of all pixels over the sphere.

Our temperature dependent analytic results reproduce  the channeled fraction of incoming Na ions as function of their energy given by the DAMA collaboration~\cite{Bernabei:2007hw}  (which does not include temperature effects), even with dechanneling effects included (shown in Fig.~2). We included only the dechanneling due to the first interaction with Tl doping atoms. A naive geometrical blocking brings our fractions to zero at very small energies in Fig.~2. In these calculations, however, the fact that the recoiling nuclei originate in lattice sites (and are not incident upon the crystal) thus blocking effects  are important, was  not properly  taken into account.
\begin{figure}[h]
\begin{minipage}{17pc}
\vspace{-1pc}
\includegraphics[width=13pc]{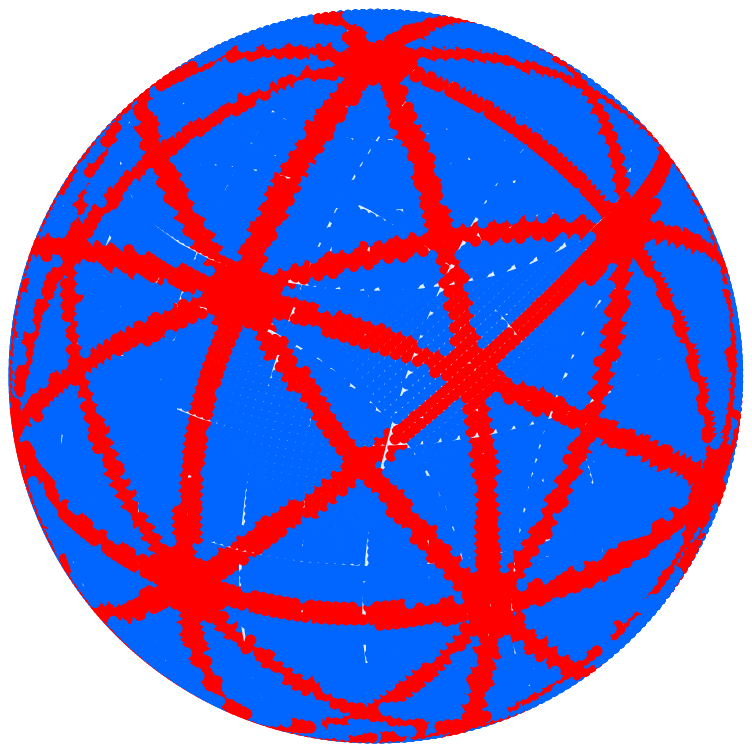}
\vspace{-2pc}
\caption{\label{label} HEALPix map of  directions of 50 keV Na ions incident upon a NaI crystal showing in red those channeled. The axial and planar channels are clearly seen.}
\end{minipage} \hspace{2pc}%
\begin{minipage}{19pc}
\includegraphics[width=17pc]{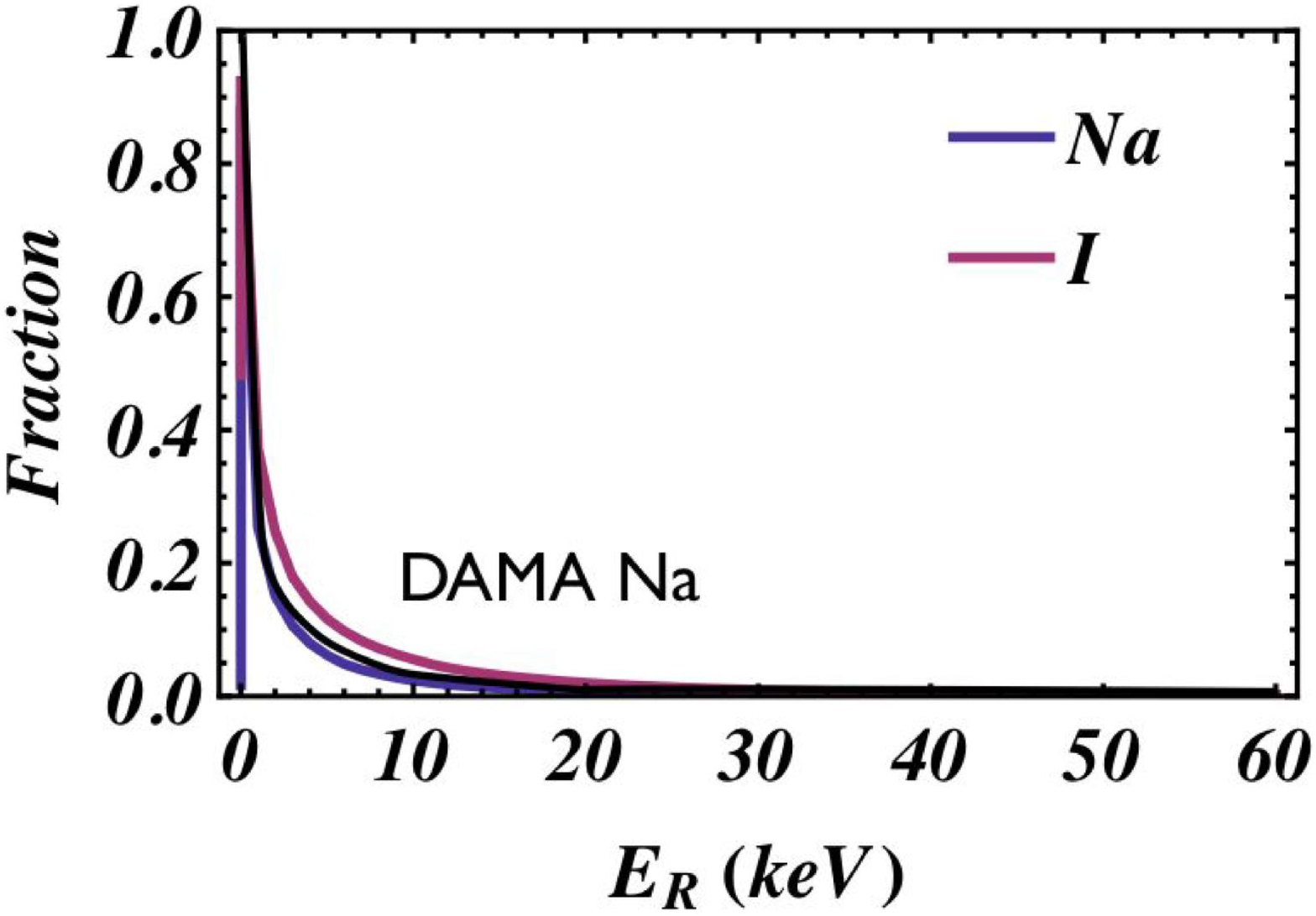}
\caption{\label{label} Preliminary results for the fraction of channeled Na and I  ions incident upon
a NaI crystal, including dechanneling, compared with the DAMA estimate (in black). }
\end{minipage}
\end{figure}
\begin{figure}[h]
\begin{minipage}{17pc}
\includegraphics[width=15pc]{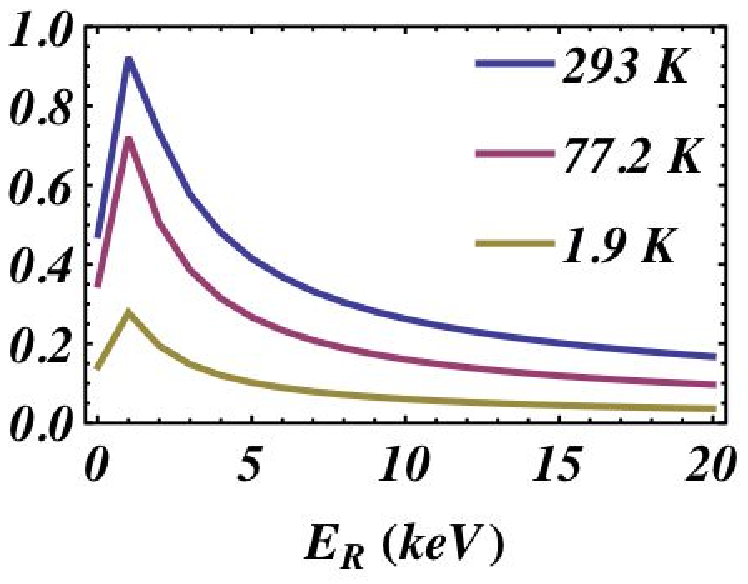}
\caption{\label{label} Preliminary results for the fraction of channeled recoiling Na nuclei in NaI (no including dechanneling). The effect is strongly temperature dependent.}
\end{minipage}\hspace{2pc}%
\begin{minipage}{17pc}
\includegraphics[width=15pc]{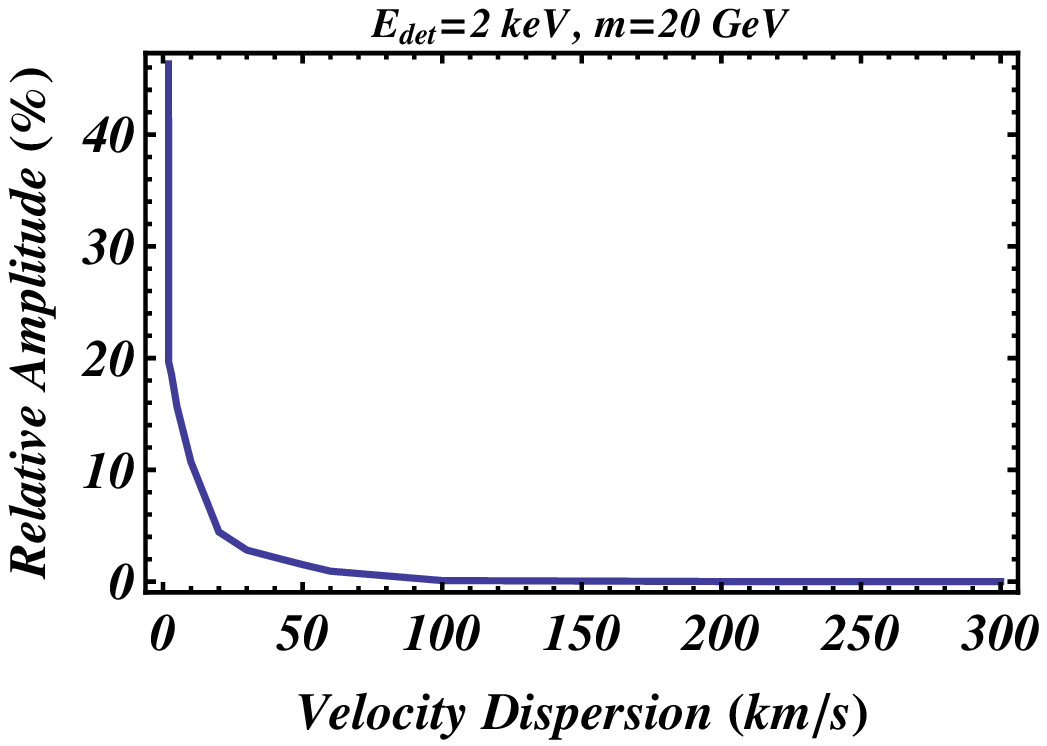}
\caption{\label{label} An example of the dependence of the daily modulation amplitude due to channeling as function of the WIMP velocity dispersion in the standard halo (changed to arbitrary unphysical values).}
\end{minipage} 
\end{figure}

As argued originally by Lindhard~\cite{Lindhard:1965}, in a perfect lattice and in the absence of energy-loss processes the probability of a particle starting from a lattice site to be channeled would be zero. It is well known in statistical mechanics that the probability of particle paths related by time reversal is the same. In optics, if a source of radiation and a point of observation are interchanged, the intensity of the light measured at the new place of observation is the same as the old.  In the same way, if the probability of an incoming channeled ion to get very close to a lattice site (where it would suffer a large angle scattering) is zero (channeled ions only undergo small angle correlated scatterings), the probability of the same ion moving in the time-reversal path, i.e. starting at a nuclear site and ending going out of the crystal in a path parallel to a channel, is zero too (this is Lindhard's {\it ``Rule of Reversibility"}). However, any departure of  the actual lattice from a perfect lattice, e.g. due to vibrations of the atoms in the lattice, violates the conditions of this argument and allow for some of the recoiling lattice nuclei to be channeled.

The channeling of particles emitted at lattice sites, such as protons scattered at large angles, due to lattice vibrations was measured and already understood in the 70's~\cite{Komaki:1970, Komaki-et-al-1971}. Due to lattice vibrations the recoiling nucleus has a probability (which we take to be Gaussian) of being at  a certain distance of the row of plane of atoms. The higher the temperature the higher the probability that the recoil starts outside a lattice site.  We use the method of Komaki and Fujimoto~\cite{Komaki:1970} to compute the fraction of channeled recoils. They use the  conservation of transverse energy  of the recoiling ion with initial energy $E_R$,  $E_\bot=E \sin^2 \psi+U= E_R \sin^2 \psi_R + U_{\rm initial}$,  valid in the continuum approximation. Here
 $\psi$ is the angle between the tangent of the trajectory and the crystal axis and $\psi_R$ is the recoil momentum after the collision.  They distinguish two types of trajectories: those which can pass over the potential barrier  $U_{max}$ separating two parallel channels in the crystal (thus particles would not be trapped within one channel) and those which cannot pass over the barrier and are, thus, channeled. Using this approach, and taking $U_{max}$  to be the value of the potential at  the Thomas-Fermi screening distance,
 we preliminary find the fractions of channeled ions shown in Fig.~3, which clearly increase with increasing temperature.

Finally, we find that the amplitude of the daily modulation of the signal depends strongly on the velocity dispersion of WIMPs (see Fig.~4) and for the standard halo model is smaller than 0.1\%. However it may be much larger for possible halo components with a smaller velocity dispersion, such as a thick disk.

{\it Acknowledgments}
This work was supported in part by the US Department of Energy Grant
DE-FG03-91ER40662, Task C. I thank S. Nussinov and F.  Avignone for useful discussions.

\end{document}